\documentstyle[12pt]{article}
\newcommand{\be}{\begin{equation}}
\newcommand{\ee}{\end{equation}}
\newcommand{\bea}{\begin{eqnarray}}
\newcommand{\eea}{\end{eqnarray}}
\newcommand{\ba}{\begin{array}}
\newcommand{\ea}{\end{array}}
\newcommand{\beas}{\begin{eqnarray*}}
\newcommand{\eeas}{\end{eqnarray*}}
\newcommand{\bes}{\begin{equation*}}
\newcommand{\ees}{\end{equation*}}

\newcommand{\nn}{\nonumber}
\newcommand{\lf}{\left}
\newcommand{\ri}{\right}
\newcommand{\f}{\frac}

\def\cA{{\cal A}}

\def\tr           {\mbox{\rm tr}\,}

\def\i2           {\mbox{$\frac{i}{2}$}}

\def\al           {\alpha}

\def\et           {\eta}

\def\ga           {\gamma}

\def\rh           {\rho}

\def\si           {\sigma}
\def\Si           {\Sigma}

\begin{document}

\title{\bf Dirac Operator on Fuzzy $AdS_2$}
\author{H. Fakhri $^{a, c}$ \thanks{Email: hfakhri@theory.ipm.ac.ir} and 
A. Imaanpur $^{b, c}$ \thanks{Email: aimaanpu@theory.ipm.ac.ir}\\ 
$^a$ {\small {\em Department of Theoretical Physics and Astrophysics,}}\\
{\small {\em Physics Faculty, Tabriz University, 
P.O.Box 51664, Tabriz, Iran}}\\
$^b$ {\small {\em Department of Physics, School of Sciences,}}\\
{\small {\em Tarbiat Modares University, P.O.Box 14155-4838,
Tehran, Iran}}\\
$^c$ {\small {\em Institute for Studies in Theoretical Physics and
 Mathematics (IPM)}} \\
{\small {\em P.O.Box 19395-5531, Tehran, Iran}}}

\maketitle

\begin{abstract}
In this article we construct the chirality and 
Dirac operators on fuzzy $AdS_2$.     
We also derive the discrete spectrum of the Dirac operator which  is 
important in the study of the spectral triple associated to $AdS_2$. It is shown that 
the degeneracy of the spectrum present in the commutative $AdS_2$ is lifted 
in the noncommutative case. The way we construct the 
chirality operator is suggestive of how to introduce the projector operators 
of the corresponding projective modules on this space. 
\end{abstract}

\section{Introduction}
Recently there has been much interest in applying the  Connes spectral triple approach to 
the study of  noncommutative manifolds such as fuzzy sphere and noncommutative $S^4$ 
\cite{BAE, CON2}. 
In this approach one starts with an algebra $\cal A$ (say the algebra of functions 
on the underlying manifold $X$), a Hilbert space $\cal H$ which is a representation of $\cal A$, 
and a Dirac operator $D$ which acts on $\cal H$. Having the triple 
$({\cal A}, {\cal H}, D)$, the geometry and topology of $X$ can be determined \cite{CON}. 
For example, the distance formula is given in terms of $D$, and knowing the spectrum 
of $D$, allows one to do the integration over $X$.   

Fuzzy sphere is the simplest example of a curved 
noncommutative manifold for which the 
spectral triple has been explicitly worked out \cite{ WATA1, WATA2}. 
Another two dimensional and non-flat example that comes to mind is the fuzzy  (noncommutative) $AdS_2$.  
Fuzzy $AdS_2$ naturally appears  
in the study of a variant form of (noncommutative) AdS/CFT 
correspondence in two dimensions \cite{HO}. 
Note that the isometry group of $AdS_2$ is $SO(2,1)$, 
and to get the fuzzy version of $AdS_2$, one 
promotes the coordinates defining the embedding of the surface in $R^{2+1}$ 
from ordinary functions to the generators of the group $SU(1,1)$. The embedding and the 
algebra satisfied by the coordinates are invariant under the automorphism group $SO(2,1)$  of the algebra and hence the name 
fuzzy $AdS_2$. 

For fuzzy sphere, a knowledge of the Dirac operator and its spectrum has allowed 
the authors of \cite{BAE} to 
compute the Chern numbers of the associated 
projective module ${\cal M}$ on fuzzy sphere.  
Here we will follow  \cite{WATA1, WATA2} to construct the Dirac operator on fuzzy 
$AdS_2$, and derive its discrete spectrum. 
In particular, we will find that the discrete spectrum of $D$ is nondegenerate. This should be contrasted with the commutative $AdS_2$ case where the discrete spectrum of $D$ is degenerate.  By this, we provide the spectral triple of fuzzy 
$AdS_2$. One may wish to build modules on this space, and, for instance, study 
the Yang-Mills theory on it. We introduce one such projective module by constructing its 
corresponding projector on the fuzzy $AdS_2$. To compute the Chern number of these 
modules, however, one has to adapt the Connes formula to the case of noncompact algebras. 

In the context of supersymmetric quantum mechanics on $AdS_2$, 
in \cite{FAKH} the spectrum of the Dirac operator, and the corresponding spinors for 
a charged particle with 
spin $\frac{1}{2}$ -- in the presence of a magnetic monopole --   
was successfully derived. There it was shown that the Dirac
spinors represent an ${\cal N}=1$ chiral supersymmetry algebra and a unitary 
parasupersymmetry algebra of an arbitrary order. Along the lines of \cite{POLY,BURES,TRA},  
it is interesting to see how 
these results are modified on fuzzy $AdS_2$. This is one of our 
motivations to study the Dirac operator on fuzzy $AdS_2$.

\section{$\Si_3$-pseudo-Hermiticity and $AdS_2$ } 

In this section we study a special representation for the Lie algebra $su(1,1)$. 
This we choose to be a non-unitary representation given by $2\times 2$ matrices. 
In this representation, the 
$\Si_3$-pseudo-Hermiticity -- which will be defined shortly -- structure of the 
algebra is very clear. 
Then we will introduce an appropriate inner product between the elements of 
the Hilbert space on which $su(1,1)$ acts. This will induce the $\Si_3$-pseudo-Hermiticity 
structure of the algebra on to the Hilbert space \cite{MOST}. 
The $\Si_3$-pseudo-Hermiticity concept will become important when we come to 
discuss the spectrum of the Dirac operator on fuzzy $AdS_2$.
   
For every irreducible unitary finite dimensional representation of
the compact semisimple Lie group $SU(2)$, one can analytically continue 
the representation to a finite dimensional necessarily non-unitary representation of
the noncompact semisimple Lie group $SU(1,1)\cong SL(2,I\!\!R)$. Both Lie groups $SU(2)$ and $SU(1,1)$ have a common maximal compact subgroup
$U(1)$, and a common complexification which is the 
Lie group $SL(2, {\bf C})$. 
Considering the lowest dimensional irreducible
representation of the Lie algebra $su(2)$, which is represented by the
Pauli matrices $\si_1$, $\si_2$ and $\si_3$, one can construct a two
dimensional non-unitary representation of the Lie algebra $su(1,1)$
\be 
\Si_1 =i \si_1\hspace{10mm}
\Si_2=i \si_2 \hspace{10mm} \Si_3=\si_3 \,\, . \label{SSHH}
\ee 
It is straightforward to show that 
\be \Si_i
\Si_j=-\eta_{ij}I + i C_{ij}^{\ \ k} \Si_k \, , \label{SS} 
\ee 
where the
indices are raised and lowered by $\et_{ij}= (1,1,-1)$. The
structure constants $C_{ijk}$'s are completely anti-symmetric in
their indices, and our convention is $C_{123}=1$. 
If we parametrize the Lie group $SU(1,1)$  of $2\times 2$ matrices
of unit determinant as $U=exp(\frac{i}{2}\theta^i \Si_i)$, they will 
satisfy the following $\Si_3$-pseudo-unitary 
relation 
\be 
U^{\dagger}\Si_3U=U\Si_3U^{\dagger}=\Si_3\, \, \label{PSOU}. 
\ee
In the next section, we will see how this property of $\Si_3$-pseudo-Hermiticity 
gets extended to the Dirac operator on fuzzy $AdS_2$.
The relation (\ref{PSOU}) induces essentially the property of
$\Si_3$-self-adjoint on $2 \times 2$ representation of the Lie algebra $su(1,1)$
\be
\Si_i^{\dagger}=\Si_3 \Si_i \Si_3 \,\,  \label{PSOH}\, ,
\ee 
where we note that the operator $\Si_3$ is self-adjoint, involutory and unitary 
\be 
\Si_3^{\dagger}=\Si_3
\hspace{10mm} \Si_3^{2}=I \hspace{10mm}
\Si_3^{-1}=\Si_3^{\dagger}\, \, \, . 
\ee 
Therefore, the Lie algebra space
$su(1,1)$ is said to be pseudo-Hermitian with respect to $\Si_3$ (or
equivalently, we say $su(1,1)$ is $\Si_3$-pseudo-Hermitian). The eigenvalues of
$\Si_3$-pseudo-Hermitian operators are known to be either real or
appear  in complex-conjugates pairs. 
Let the linear operator $u:{\cal H}\rightarrow {\cal H}$ be an arbitrary element
of the Lie algebra $su(1,1)$ acting on a separable Hilbert space
${\cal H}$. $u$ is expressed as a linear combination of the
traceless $2\times 2$ matrices $\Si_1$, $\Si_2$ and $\Si_3$ with
real determinants. 
If we define the $\Si_3$-adjoint for
every $\Psi$ belonging to the Hilbert space ${\cal H}$ as 
\be
\overline{\Psi}=\Psi^{\dagger}\Si_3\, \, , 
\ee 
then we can introduce a 
natural indefinite inner product ``$\ast$" of the two arbitrary elements
$\Psi$ and $\Phi$ belonging to the Hilbert space ${\cal H}$, as 
\be
\Psi\ast\Phi=\overline{\Psi}\Phi\, \, . 
\ee 
This  has the
following property for every arbitrary element $u$ of the Lie algebra $su(1,1)$ 
\be
\Psi\ast \left(u\Phi\right)=
\overline{\Psi}u\Phi=\Psi^{\dagger}\Si_3u\Phi=\Psi^{\dagger}u^{\dagger}\Si_3\Phi
=\left(u\Psi\right)^{\dagger}\Si_3\Phi=
\overline{\left(u\Psi\right)}\Phi=\left(u\Psi\right)\ast \Phi \, \, , 
\ee 
which means that all the generators belonging to $su(1,1)$ are
$\Si_3$-pseudo-Hermitian with respect to the inner product ``$\ast$". 
A Hilbert space equipped with such an indefinite inner product
is called the Krein space \cite{AZIZ}. Using Eq. (\ref{SS}) the Lie and Clifford
algebras corresponding to the generators $\Si_i$ are obtained, 
respectively as  
\bea
&&[\Si_i , \Si_j]= 2i C_{ij}^{\ \ k} \Si_k  \\
&&\{\Si_i , \Si_j\}= -2 \eta_{ij} I\, \, . \label{ANTIS}
\eea
It is straightforward to show that the structure constants $C_{ij}^{\ \ k}$
satisfy the following relation 
\be
C_{im}^{\ \ \ k} \eta^{ij} C_{jl}^{\ \ n}=\eta_{m}^{\ \ n}\eta_{l}^{\ k}
- \eta_{ml} \eta^{kn}. \label{SABET} 
\ee

Next let us discuss the geometrical structure of the Euclidean $AdS_2$. 
The Euclidean $AdS_2=SU(1,1)/U(1)$
is defined through the following embedding in 3-dimensional flat Minkowskian space
 ($l>0$)
\be
{\bf x}\cdot{\bf x}:=x_i\eta^{ij}x_j=-l^2\, \, . \label{EMB}
\ee
This could also be thought of as the poincare upper half-plane $H^2=\{(x,y)\in I\!\! R^{2} : y>0\}$ with
the Riemannian metric
\be
ds^2=l^2 \frac{dx^2+dy^2}{y^2}\,\, .\label{MET}
\ee
If we make the following coordinates transformations 
\be
x=\frac{2\tanh\frac{\rh}{2} \sin\tau}{1+2\tanh\frac{\rh}{2}
\cos\tau +\tanh^2\frac{\rh}{2}} \hspace{20mm}
y=\frac{1-\tanh^2\frac{\rh}{2}}{1+2\tanh\frac{\rh}{2}
\cos\tau +\tanh^2\frac{\rh}{2}}\, ,
\ee
the metric (\ref{MET}) will transform to 
\be
ds^2 =l^2 (d\rh^2 + \sinh ^2\rh \, d\tau^2 )\,\, . \label{METRIC}
\ee
The above metric can also be obtained by simply inserting the 
embedding coordinates
\bea
&& x_1= l\sinh \rh \cos \tau \\ \nn
&& x_2 =l\sinh \rh \sin \tau \\ \nn 
&& x_3=l\cosh \rho \nn 
\eea
into the Minkowskian metric 
\bea
ds^2=dx_1^2 +dx_2^2 -dx_3^2\, .
\eea
Also note that the metric (\ref{METRIC}) has a negative scalar curvature $R=-2/l^2$.

\section{ $U(1)$ principal fiber bundles over $AdS_2$}

For the sake of completeness, here we discuss the Hopf construction of 
the $U(1)$ bundles over $AdS_2$. This will prove useful in the 
discussion of the noncommutative analogue of bundles over $AdS_2$, i.e., 
the construction of projective modules over fuzzy $AdS_2$. 
 
The $U(1)$ principal fiberation $\pi$ of the total space $AdS_3$
over $AdS_2$ can be realized as follows\footnote{For a similar construction 
on $S^2$ see \cite{LANDI}.}.  
First we define the total space 
in ${{\bf C}}^{2}$ through the embedding 
\be
AdS_3=\left\{\left(z_1,z_2\right)\in {{\bf C}}^{2}, \hspace{5mm} \left|z_1\right|^2
-\left|z_2\right|^2=-l^2\right\}\, . \label{ADS3}
\ee
In the framework of principal fiber bundle, this is projected over the base manifold 
$AdS_2$ as 
\be
U(1)\stackrel{\mbox{right}\hspace{1mm} U(1)-\mbox{action}}{\hookrightarrow} AdS_3
\stackrel{\pi}{\longrightarrow} AdS_2 \,.
\ee
The right $U(1)$-action transforms the point $\left(z_1,z_2\right)$
of $AdS_3$ onto another point of $AdS_3$ 
\begin{equation}
\begin{array}{llll}
&&\nonumber\\
&&\left.
\begin{array}{llll}
&&AdS_3 \times U(1)\rightarrow  AdS_3 \\
&&\left(z_1,z_2\right)w=\left(z_1w,z_2w\right)
\end{array}
\right\}
\rightarrow \left|z_1w\right|^2-\left|z_2w\right|^2 =\left|z_1\right|^2
-\left|z_2\right|^2=-l^2\,.
\end{array}
\end{equation}
The complex Hopf bundle projection $\pi:AdS_3\rightarrow AdS_2$ is given by 
\bea
&&x_1=\frac{1}{l}\left(z_1{\bar z}_2+z_2{\bar z}_1\right) \nonumber \\
&&x_2=\frac{i}{l}\left(z_1{\bar z}_2-z_2{\bar z}_1\right)  \nonumber \\
&&x_3=\frac{1}{l}\left(\left|z_1\right|^2+\left|z_2\right|^2\right)\, .
\label{BUNMAP}
\eea
The relation (\ref{EMB}) can now be directly checked. 
Moreover, the Hopf fibration of $U(1)$ bundles over $AdS_2$ can also be expressed 
in the group theoretical framework as follows.   
There is a canonical one-to-one correspondence between the points on $AdS_3$ and 
the elements of $SU(1,1)$, $g:AdS_3\rightarrow SU(1,1)$. 
It is sufficient to choose the one-to-one map $g$
as 
\begin{eqnarray}
g\left( \begin{array}{cc} z_1
\\ z_2
\end{array} \right)=\frac{1}{l} \left( \begin{array}{cc} {\bar z}_2 & iz_1
\\ -i{\bar z}_1  & z_2
\end{array} \right)=:U\, ,
\end{eqnarray}
in which $z_1$ and $z_2$ are arbitrary elements of $AdS_3$. It is easy to
check that it has the unit determinant, and also, satisfies $\Si_3$-pseudo-unitary
relation (\ref{PSOU}). The homogeneous manifold $SU(1,1)/U(1)$ can be obtained by
the Hopf projection as 
\begin{eqnarray}
\pi\left(U\right):=lU\Si_3U^{-1}=-{\bf x}\cdot{\bf \Si}\, , \label{PUS}
\end{eqnarray}
where the coordinates $x_i$ (for homogeneous manifold $SU(1,1)/U(1)$)
are obtained from the complex coordinates $z_1$ and $z_2$
(for the group manifold $SU(1,1)$) via Eqs. (\ref{BUNMAP}).
The right $U(1)$-action over $SU(1,1)$ keeps the base point ${\bf x}$
fixed in the sence that all the elements
$U exp\left(\frac{i}{2}\theta^3\Si_3\right)$ of $SU(1,1)$ are projected
onto the same point ${\bf x}$ on the base manifold
$SU(1,1)/U(1)$. Therefore, we can identify any equivalance class
$\left[U\right]=U exp\left(\frac{i}{2}\theta^3\Si_3\right)\in SU(1,1)/U(1)$
with the point ${\bf x}\in AdS_2$ via the projection map (\ref{PUS}).

\section{Fuzzy $AdS_2$ and finite dimensional Schwinger representation 
of the algebra ${\cal A}_N$}

As mentioned in introduction, to define the fuzzy $AdS_2$, we promote the 
coordinates $x_i$'s of $AdS_2$ to play the role of the generators of 
$su(1,1)$, in some unitary representation. For a given integer $N$ ($N>2$), 
let (${\cal A}_N$,${\cal L}_N$) denote the space of all analytic functions 
of coordinates, and derivations on fuzzy $AdS_2$, respectively. As in the 
case of fuzzy sphere we can work out a Schwinger representation for ${\cal A}_N$.   
To start with, introduce a real parameter 
$\al$ controling the strength of the noncommutativity, and take the coordinates $x_i$'s 
to satisfy the relation
\be
\left[x_i , x_j\right]= i\al C_{ij}^{\ \ k} x_k \, , \label{CMX}
\ee
such that $\f{2}{ \al}x^i$'s satisfy the commutation relations of the Lie algebra $su(1,1)$.
Like $SU(2)$, the Lie group $SU(1,1)$ is a group of rank $1$, and so possesses one invariant
Casimir operator whose eigenvalues label the irreducible representations.
The noncommutative relation (\ref{CMX}) together with the embedding (\ref{EMB}), 
which now ascribes a negative constant
value to the Casimir operator, defines the fuzzy 
$AdS_2$. In fact, this is an analytic continuation of the noncommutative structure of the fuzzy sphere to that of $AdS_2$ via the metric $\et_{ij}$.  
The passage from the fuzzy sphere to the fuzzy $AdS_2$
is equivalent analytically to the passage from the ordinary Euclidean Clifford algebra 
to the Minkowskian Clifford algebra. 

The generators $L_i$ of ${\cal L}_N$ are defined by the adjoint action of 
$x_i$ on the space ${\cal A}_N$:
\be
\frac{1}{\alpha}ad_{x_i}x_j=\frac{1}{\alpha}\left[x_i,x_j\right]=:L_ix_j\, \label{DERIV}.
\ee
This induces the commutation relations of the Lie algebra $su(1,1)$ on ${\cal L}_N$ 
\be
\left[L_i , L_j\right]= i C_{ij}^{\ \ k} L_k \, . \label{CML}
\ee
The bigger algebra (${\cal A}_N$ , ${\cal L}_N$) includes the commutation relations
(\ref{CMX}) and  (\ref{CML}), as well as
\be
\left[L_i , x_j\right]= i C_{ij}^{\ \ k} x_k \, .  \label{LX}
\ee

We are now going to introduce a version of the Schwinger operator bases which realizes a representation of the Lie algebra $su(1,1)$ using the $\Si_3$-pseudo-Hermitian matrices (\ref{SSHH}). In this formulation, one considers the elements of ${\cal A}_N$
acting on a finite dimensional Hilbert space ${\cal F}_N$ spanned by a set of 
orthonormal bases $\left\{\left|k\right\rangle\right\}_{k=0}^{k=N-2}$.
Then, the algebra ${\cal A}_N$ can be identified with the algebra of 
$(N-1)\times (N-1)$ complex matrices, which act on an $(N-1)$-dimensional Hilbert space
${\cal F}_N$. According to the discussion above, the Hilbert space ${\cal F}_N$ can be constructed by acting a pair of creation and annihilation operators ${{\bf a}^{\dagger}}_{b}$ and ${\bf a}^{b}$ 
($b=1$, $2$) on the vacuum state $\left|0\right\rangle$, i.e.,
\be
\left|k\right\rangle=\frac{1}{\sqrt{k!(N-2-k)!}}\left({{\bf a}^{\dagger}}_{1}\right)^k
\left({{\bf a}^{\dagger}}_{2}\right)^{N-2-k}\left|0\right\rangle \, \hspace{20mm}k=0,1, 
\cdots ,N-2 \, ,
\ee 
where the commutation relations are
\be
\left[{\bf a}^{a} , {\bf a}^{b}\right]=\left[{{\bf a}^{\dagger}}_{a} , 
{{\bf a}^{\dagger}}_{b}\right]=0 \hspace{20mm}
\left[{\bf a}^{a} , 
{{\bf a}^{\dagger}}_{b}\right]=\delta^a_b \, .
\ee
The number operator 
${\bf N}:={{\bf a}^{\dagger}}_{b}{\bf a}^{b}$ has the eigenvalue $N-2$ over 
the Hilbert space ${\cal F}_N$ 
\be
{\bf N}|_{_{{\cal F}_N}}=N-2 \,.
\ee

The Schwinger representation for the bases of the algebra ${\cal A}_N$ is now represented
as follows
\be
x_i=\frac{1}{2}\al \left(\Si_{i}\right)^a_{\, \, b} {{\bf a}^{\dagger}}_{a} 
{\bf a}^{b}\,. \label{SCH}
\ee
It is straightforward to see that the generators (\ref{SCH}) satisfy the algebra of fuzzy  $AdS_2$ (\ref{CMX}), as well as the following commutation relations:
\bea
\left[x_i , {{\bf a}^{\dagger}}_{a}\right]=\frac{1}{2}\al \left(\Si_{i}\right)^b_{\, \, a} {{\bf a}^{\dagger}}_{b} \nonumber \\
\left[x_i , {\bf a}^{a}\right]=\frac{-1}{2}\al \left(\Si_{i}\right)^a_{\,\, b} 
{\bf a}^{b}\, .
\eea
The Casimir operator of the generators (\ref{SCH}) is  
\be
{\bf x}\cdot{\bf x}=\frac{-\al^2}{4} {\bf N}({\bf N}+2) \,. \label{CASIMIR}
\ee
Restricting the equations over the Hilbert space ${\cal F}_N$, and 
comparing Eq. (\ref{EMB}) with (\ref{CASIMIR}), we conclude that
\be
\alpha=\frac{2l}{\sqrt{N(N-2)}} \, , \label{CONS}
\ee
which implies that 
the commutative limit $(\alpha\rightarrow 0)$ corresponds to the limit 
$N\rightarrow \infty$.

\section{Chirality and Dirac operators}

Our aim in this section is to construct the Dirac operator on 
fuzzy $AdS_2$. This is the most important ingredient in the 
Connes construction of noncommutative manifolds.  
To define the Dirac operator, however, first we need to introduce the 
chirality operator $\ga$. The existence of this latter operator provides a 
$Z_2$ grading of the Hilbert space ${\cal H}$. $\ga$ has the following properties:

\vspace{3mm} 

$\bullet$ Commutes with the elements of the algebra ${\cal A}_N$, and squares to one.

\vspace{3mm}

$\bullet$ Has a standard commutative limit.
\vspace{3mm}
\newline
This chirality operator, however, instead of being Hermitian as in 
the case of 
compact algebras, turns out to be $\Si_3$-pseudo-Hermitian on fuzzy $AdS_2$.
The Dirac operator $D$ is then constructed such that: 

\vspace{3mm}

$\bullet$ It anticommutes with $\ga$. 

\vspace{3mm}

$\bullet$ Reduces to the conventional Dirac operator on commutative $AdS_2$ \cite{TRA}, when 
the noncommutativity parameter $\al$ is sent to zero. 
\vspace{3mm}\newline
Notice that the above requirements on $\ga$ and $D$ do not uniquely fix the 
operators. 
Let us represent the Dirac and chirality operators by the $2$-component spinors 
$\Psi=\left( \begin{array}{c}\psi_1 \\ \psi_2 \end{array}\right)$ as the elements of 
${\cal A}_N$-bimodule ${\cal H}:={{\bf C}}^{2}\otimes {\cal A}_N$. 

Following \cite{WATA2}, to construct $\ga$, we introduce the
opposite algebra ${\cal A}_N^0$ with the product rule between the elements
\be
x_i^0 x_j^0 \equiv (x_j x_i)^0 \, .
\ee
In other words, the generators $x_i^0$, which form the opposite algebra ${\cal A}_N^0$, 
are acting from 
the right hand side on ${\cal A}_N$-bimodule:
\be
x_i^0 \Psi=\Psi x_i \hspace{20mm} \forall \Psi\in {\cal H}. \label{ZEROD}
\ee 
From these, we obtain the commutation relation of this algebra to be
\be
\left[x_i^0 , x_j^0\right]= -i\al C_{ij}^{\ \ k} x_k^0 \, . \label{ZERO}
\ee
Now let us define the chirality operator to be
\be
\ga =\f{1}{{\cal N}_N}(\Si_i \et^{ij}x_j^0 +\f{\al}{2}I)\, ,
\ee
where 
\be
{\cal N}_N =\sqrt{l^2+\frac{\al^2}{4}}  
=\frac{l(N-1)}{\sqrt{N(N-2)}}\,. \nn 
\ee
It is easy to check that $\ga$ is involutory, just note that
\bea
(\Si_i \et^{ij}x_j^0)(\Si_k \et^{kl}x_l^0) &=& 
\et^{ij}\et^{kl}(-\et_{ik}I +iC_{ik}^{\ \ n}\Si_n)x^0_jx^0_l \nn \\
&=& -\et^{jl}x^0_jx^0_l I +iC^{jln}\Si_nx^0_jx^0_l \nn \\
&=& l^2I +\f{\al}{2}C^{jln}C_{jl}^{\ \ r}\Si_nx^0_r \nn \\
&=& l^2I -\al\et^{nr}\Si_nx^0_r \, ,
\eea  
so 
\be
\ga^2 = \f{1}{{\cal N}_N^2}\lf(\f{\al^2}{4}I +\al\Si_i\et^{ij}x^0_j +l^2I 
-\al\Si_i\et^{ij}x^0_j\ri)=I\, .\label{GAMMA2}
\ee
Note that the chirality operator $\ga$ is constructed so 
that it commutes with the elements of
algebra ${\cal A}_N$. Moreover, since $\Si_i^\dag = \Si_3 \Si_i \Si_3$, 
it is $\Si_3$-pseudo-Hermitian
\be
\ga^{\dagger}=\Si_3 \ga \Si_3 \,. \label{PSHER}
\ee

Next we introduce the Dirac operator
\be
D=\f{-i}{l\al}\ga\, C_{ij}^{\ \ k}\et^{il}\et^{jm}\Si_lx^0_m x_k\, . \label{DIRAC1}
\ee
In the following we show that $\ga$ anticommutes with $D$. 
First note that
\bea
(\et^{ij}\Si_ix_j^0)(\Si_lx_k^0)=\et^{ij}\Si_i\Si_l x_j^0x_k^0 &=& 
\et^{ij}(-\Si_l\Si_i -2\et_{il}I)(x^0_kx^0_j -i
\al C_{jk}^{\ \ r}x_r^0) \nn \\
&=& -(\Si_lx^0_k)(\et^{ij}\Si_ix^0_j) +i\al \et^{ij}C_{jk}^{\ \ 
r}\Si_l\Si_ix^0_r 
-2x^0_lx^0_k I \, ,
\eea 
where in the first line we have used Eq. (\ref{ANTIS}), and 
the relation 
(\ref{ZERO}). So we learn that
\bea
\lf\{ C_{ij}^{\ \ k}\et^{il}\et^{jm}\Si_l x_m^0 \, , \et^{ij}\Si_ix^0_j 
\ri\} &=& 
i\al C_{ij}^{\ \ k}C_{sm}^{\ \ p}\et^{il}\et^{jm}\et^{rs}\Si_l\Si_r 
x^0_p 
-2 C_{ij}^{\ \ k}\et^{il}\et^{jm}x^0_lx^0_m I\nn \\
&=& -\al  C_{ij}^{\ \ k}\et^{il}C_{lr}^{\ \ n}\et^{jm}\et^{rs}C_{sm}^{\ 
\ p} \Si_n x^0_p\nn \\
&=& -\al \et^{nm}\et^{ks} C_{sm}^{\ \ p} \Si_nx^0_p \nn \\
&=& -\al \et^{ni}\et^{pj}C_{ij}^{\ \ k}\Si_n x^0_p \, .
\eea
The second and the third equality follow from Eqs. (\ref{SS}) and (\ref{SABET}). 
Since $x_i$ and $x_j^0$ commute with each other, we conclude that
\bea
\{D\, , \ga \} &=& \f{-i}{l\al {\cal N}_N}\ga 
\lf\{C_{ij}^{\ \ k}\et^{il}\et^{jm}\Si_l x_m^0 x_k\, , 
\et^{ij}\Si_ix^0_j \ri\} +
\f{\al}{{\cal N}_N}D \nn \\
&=& \f{i}{l{\cal N}_N}\ga C_{ij}^{\ \ k}\et^{il}\et^{jm}\Si_lx^0_m 
x_k+\f{\al}{{\cal N}_N} D\nn \\
&=& 0 \, .\label{ADIRAC}
\eea
Further, Eq. (\ref{PSHER}) implies that
\be
D^\dag = \Si_3 D \Si_3 \label{PDIRAC}\, , 
\ee
so it is $\Si_3$-pseudo-Hermitian.

To proceed further and to discuss the commutative limit of the Dirac operator, 
there are some definitions and algebra among the operators in $({\cal A}_N , {\cal L}_N)$ that should be discussed.  
The commutation relations (\ref{CML}), (\ref{ZERO}) and also 
\be
\left[L_i , x_j^0\right]= i C_{ij}^{\ \ k} x_k^0 \,  \label{LZERO}
\ee
which is obtained from Eqs. (\ref{DERIV}) and (\ref{ZEROD}), define the algebra  
$\left({\cal A}_N^0 , {\cal L}_N\right)$ for the fuzzy $AdS_2$.
Note that the generators of the algebra ${\cal L}_N$ can also be written as a linear combination of the generators of the algebras ${\cal A}_N$ and ${\cal A}_N^0$ as 
\be
L_i = \alpha^{-1}\left(x_i - x_i^0\right). \label{LXX}
\ee
Introduce the following operators in  
$\left({\cal A}_N , {\cal L}_N\right)\otimes M_2({\bf C})$ 
\bea
&&\chi:=x_i \eta^{ij} \Si_j \\
&&\Lambda:=L_i \eta^{ij} \Si_j \\
&&\Sigma:=-i C_{ij}^{\ \ k} \eta^{il} \eta^{jm}x_l L_m \Si_k \, ,
\eea
which are invariant under the $SU(1,1)$ transformations.\footnote{Using the commutation relations of the algebra $({\cal A}_N , {\cal L}_N)$,  
the (anti)commutation relations of the operators $\chi$,
$\Lambda$ and $\Sigma$ are 
\bea
\frac{1}{2}\left\{\chi , \chi\right\}&=&l^2+\alpha \chi \nonumber \\    
\frac{1}{2}\left\{\Lambda ,\Lambda\right\}&=&-{\bf L}\cdot{\bf L}+ \Lambda \nonumber \\
\frac{1}{2}\left\{\Si , \Si\right\}&=&-\alpha {\bf x}\cdot{\bf L}+l^2{\bf L}\cdot{\bf L}+\left({\bf x}\cdot{\bf L}\right)^2 
-\alpha \Si +\alpha {\bf x}\cdot{\bf L} \Lambda +l^2\Lambda\, , 
\nonumber \\ & & \nonumber \\ 
\left\{\chi , \Lambda\right\}&=&2\left(\chi-{\bf x}\cdot{\bf L}\right)  \nonumber \\
\left[\chi , \Lambda\right]&=&-2\left(\chi+\Si \right)\, , \nonumber \\ & & \nonumber \\
\left\{\Si , \Lambda\right\}&=&2\left(\Si+{\bf x}\cdot{\bf L}\right)  \nonumber \\
\left[\Si , \Lambda\right]&=&2\left(\chi {\bf L}\cdot{\bf L}-\Lambda {\bf x}\cdot{\bf L} \right)\, , \nonumber \\ & & \nonumber \\
\left\{\Si , \chi\right\}&=& -2 l^2 +\alpha \left(\Si-\chi\right)  \nonumber \\  
\left[\Si , \chi\right]&=& -2 l^2 \left(I-\Lambda\right)-2 \alpha {\bf x}\cdot{\bf L}
+\left\{\chi , {\bf x}\cdot{\bf L}\right\}. \nonumber
\eea} 
Using Eq. (\ref{PSOH}), one can easily show that 
\bea
\chi^{\dagger}&=&\Si_3 \chi \Si_3     \label{PSX} \\
\Lambda^{\dagger}&=&\Si_3 \Lambda \Si_3     \label{PLX}\\
\Sigma^{\dagger}&=&-\Si_3\left(\Sigma+2\chi \right)  \Si_3 \, .  \label{PSIX}
\eea
 
If we define\footnote{Note that the generators 
$J_i$ are $\Si_3$-pseudo-Hermitian with respect 
to the inner product ``$\ast$" defined in the ${\cal A}_N$-bimodule 
${\cal H}:={{\bf C}}^{2}\otimes {\cal A}_N$, because
\[
\Psi\ast \left(J_i\Phi\right)=
\Psi^{\dagger}\Si_3\left(L_i+\frac{1}{2}\Si_i\right)\Phi=\Psi^{\dagger}
\left(L_i+\frac{1}{2}\Si_i^{\dagger}\right)\Si_3\Phi
=\left(\left(L_i+\frac{1}{2}\Si_i\right)\Psi\right)^{\dagger}\Si_3\Phi=
\left(J_i\Psi\right)\ast \Phi \, \,. \nn
\] 
Also note that the generators 
$J_i$, which like the bases of  
${\cal A}_N$ and ${\cal L}_N$, constitute the Lie algebra $su(1,1)$, satisfy the 
following commutation relations
\[
\left[L_i , J_j\right]= i C_{ij}^{\ \ k} L_k \nn \, \, . 
\]}
\be
J_i=L_i+\frac{1}{2} \Si_i \label{JLS} \, ,
\ee
and
\be
\Omega_i=C_{ij}^{\ \ k} \eta^{jl} x_l \Si_k 
\ee
then, using Eqs. (\ref{SS}), (\ref{SABET}) and (\ref{CMX}), it follows that
\be
C_{ij}^{\ \ k} \eta^{il} \eta^{jm} \Si_l x_m^0 x_k= - \alpha {\bf \Omega}\cdot{\bf J}
=i \alpha \left(\Si + \chi \right) \, .  \label{OJSX}
\ee
Therefore the Dirac operator 
can be written in the following compact forms
\be
D=\frac{i}{l} \ga {\bf \Omega}\cdot{\bf J}=\frac{1}{l} \ga 
\left(\Si + \chi \right) \, .\label{DOJ}
\ee
In this form the $\Si_3$-pseudo-Hermiticity relation (\ref{PDIRAC}) can easily be 
derived from the second equality in (\ref{DOJ}) upon taking its dagger and using Eqs. 
(\ref{GAMMA2}), (\ref{PSHER}), (\ref{ADIRAC}), (\ref{PSX}) and (\ref{PSIX}). 

Now let us discuss the status of the Casimir operators in ${\cal L}_N$. Unlike 
the commutative case where the only Casimir is ${\bf L}\cdot{\bf L}$, here we get 
an additional Casimir operator. Though, it will  
scale to zero in the commutative limit $\al\to 0$. 
First of all, it is clear that the operator 
${\bf L}\cdot{\bf L}$, the Casimir operator of the 
bases of the algebra ${\cal L}_N$, satisfies the following relation 
\be
\left[L_i , {\bf L}\cdot{\bf L}\right]=0 \, .\label{LLL}
\ee
On the other hand, from the commutation relations (\ref{CML})
and (\ref{LX}), one obtains
\be
\left[L_i , {\bf x}\cdot{\bf L}\right]=0  \, ,   \label{LXL}
\ee
giving the second Casimir mentioned above. If we define 
the generalized momentum operators
\be
P_i=-i C_{ij}^{\ \ k} \eta^{jl} x_l L_k \, ,   \label{PPP}
\ee
which are invariant under the $SU(1,1)$ group transformations, 
we can obtain the following relations in the algebra 
$({\cal A}_N , {\cal L}_N)$ 
\bea
&&\left[x_i , {\bf L}\cdot{\bf L}\right]=2 \left(x_i+ P_i\right)  \label{XLLP} \\
&&\left[x_i , {\bf x}\cdot{\bf L}\right]=\alpha \left(x_i+ P_i\right)\,.  \label{XXLP}
\eea
And finally, one can obtain
the action of both operators ${\bf L}\cdot{\bf L}$ and ${\bf x}\cdot{\bf L}$
on the polynomials in the 
algebra ${\cal A}_N$ using the inductive procedure $(n=1,2,3,\cdots)$
\bea
&&{\bf L}\cdot{\bf L}\left(x_i\right)^n= -2 n \left(x_i\right)^n    \label{LL}  \\
&&{\bf x}\cdot{\bf L}\left(x_i\right)^n= - n \alpha \left(x_i\right)^n \,.  \label{XL}
\eea 
Comparing Eqs. (\ref{LLL}) and (\ref{LXL}), (\ref{XLLP}) and
(\ref{XXLP}), and also, (\ref{LL})  and (\ref{XL}), it is readily
found that
\be
{\bf x}\cdot{\bf L}=\frac{\alpha}{2} {\bf L}\cdot{\bf L} \, ,   \label{XLLL}
\ee
which shows that the commutative limit is reached when $\alpha\rightarrow 0$. 
 
We notice that in the commutative limit $\alpha\rightarrow 0$ 
( or $N\rightarrow \infty$ ), the generators of the opposite algebra, $x_i^0$'s, 
get transformed into the generators of ${\cal A}_N$, i.e. 
$x_i$'s. Therefore, using Eqs. (\ref{SS}) and (\ref{SABET}) we get
\be
D_{\infty}=-\left(\Lambda-I\right)-\frac{1}{l^2}\chi {\bf x}\cdot{\bf L}\, . \label{DINFTY}
\ee
However, taking into account the relation (\ref{XLLL}), the second term of $D_{\infty}$ also vanishes, and we are left with
\be
D_{\infty}=-\left(\Lambda-I\right)\, , \label{DINFT}
\ee
which is in agreement with the results of \cite{TRA}.
Therefore Dirac operator (\ref{DIRAC1}) 
reduces to the conventional Dirac operator on
commutative $AdS_2$ when
the noncommutativity parameter $\al$ is sent to zero.

Before concluding this section, let us briefly discuss the related subject of 
projective modules on fuzzy $AdS_2$. Projective modules are analogue of fiber 
bundles on noncommutative spaces. So if we are to study, for instance, the Yang-Mills 
theory on noncommutative spaces, the concept of projective modules becomes 
indispensable. 
To construct a projective module, of a rank say 2, we proceed as follows. 
First we introduce a projector $p$, which is a 2$\times$2 matrix 
with elements in the algebra ${\cal A}$. This, in turn, allows us to 
define the projective 
module ${\cal M}= p\, {\cal A}^2 $ of sections $s\in \cA ^2$ such that $p\circ s =s$. 
The connection 
$\nabla = p\circ d$ on ${\cal M}$ 
is then a map from sections to one-form valued sections in ${\cal M}$. 

From what we did in the case of chirality operator $\ga$, which squares 
to $I$, it is not difficult to guess the form of $p$. We define
\be
p=\f{1}{2}(I+\Pi)\, ,\label{P}
\ee
where
\be
\Pi =\f{1}{{\cal N}_N}({\widetilde \Si}_i \et^{ij}x_j-\f{\al}{2}I)\, ,
\ee
such that $\Pi^2 =I$, therefore $p^2=p$ and $p$ is a projector.
Note that in the above, ${\widetilde \Si}_i$'s are again the 2-dimensional 
representations of the $su(1,1)$ generators such that 
$[{\widetilde \Si}_i , \Si_j]=0$. Now, as mentioned earlier, $p$ 
defines the projective module ${\cal M}= p\, {\cal A}^2$ over ${\cal A}$ with 
the sections $s\in \cA ^2$ such that $p\circ s =s$.
Let us introduce the differential operator $d$ through
\be
du=i[D , u]\, ,
\ee 
for any $u$ in the Lie algeba of $su(1,1)$. If $s$ is a section of the 
projective module ${\cal M}$ then $s=p\circ s$, and we define the connection 
on ${\cal M}$ to be $\nabla =d\circ p$. Therefore the curvature of $\cal M$ is 
$\nabla^2= p dp^2$ with the first Chern class of 
\be
C_1(p)=\tr(pdp^2)\, .\label{C}
\ee
For compact algebras, the first Chern number of ${\cal M}$ is then calculated by 
taking the Dixmier trace over the Hilbert space ${\cal H}$ \cite{CON}. 
For the projective module defined by the 
projector (\ref{P}), it is straightforward to compute the first Chern class 
defined in (\ref{C}). But to calculate the first Chern number and to see whether 
the associated module is nontrivial needs a modification of the Connes formula 
 to the case of noncompact algebras. This is a problem that we would like 
to further study in future works.

\section{Discrete spectrum of Dirac operator on fuzzy $AdS_2$}

In order to analyze the spectrum of the Dirac operator on fuzzy $AdS_2$,
first note that using Eqs. (\ref{EMB}) and (\ref{CONS}) we can write 
\be
{\bf X}\cdot{\bf X}={\bf X^0}\cdot{\bf X^0}=\frac{-N}{2}\left(\frac{N}{2}-1\right)\, ,   \label{XXXX}
\ee
where 
\be
X_i=\alpha^{-1} x_i \hspace{20mm} X_i^0=-\alpha^{-1} x_i^0 \,. \label{XAX}
\ee
The square of the Dirac operator, on the other hand, is calculated by using 
Eqs. (\ref{SS}), (\ref{SABET}), (\ref{CMX}), (\ref{ZERO}), and the definition 
(\ref{DIRAC1}) 
\be
\frac{l^2}{\alpha^2}D^2={\bf X}\cdot{\bf X}\,{\bf X^0}\cdot{\bf X^0} 
-{\bf X}\cdot{\bf X^0}\left({\bf X}\cdot{\bf X^0}-I+{\bf \Si}\cdot\left({\bf X}+{\bf X^0}\right)\right).   \label{DXXXX}
\ee
This can be simplified further using (\ref{LXX}) and (\ref{JLS})
\bea
&&{\bf X}\cdot{\bf X^0}=\frac{1}{2}\left({\bf L}\cdot{\bf L}-{\bf X}\cdot{\bf X}-{\bf X^0}\cdot{\bf X^0}\right)  \label{XXSQ}   \\
&&{\bf \Si}\cdot\left({\bf X}+{\bf X^0}\right)={\bf J}\cdot{\bf J}-{\bf L}\cdot{\bf L}
+\frac{3}{4}I \,,  \label{SXXJJ}
\eea
the last equation is obtained noticing 
$\left(\frac{1}{2}{\bf \Si}\right)\cdot\left(\frac{1}{2}{\bf \Si}\right)
=-s(s-1)I$ with $s=\frac{-1}{2}$.
Finally, applying Eqs. (\ref{XXSQ}) and (\ref{SXXJJ}), the square of Dirac operator 
transforms to the form
\bea
\frac{l^2}{\alpha^2}D^2&=&{\bf X}\cdot{\bf X}\,{\bf X^0}\cdot{\bf X^0} 
-\frac{1}{2}\left({\bf L}\cdot{\bf L}-{\bf X}\cdot{\bf X}-{\bf X^0}\cdot{\bf X^0}\right)
\nonumber\\
&&\times \left[\frac{1}{2}\left({\bf L}\cdot{\bf L}-{\bf X}\cdot{\bf X}-{\bf X^0}\cdot{\bf X^0}\right) 
+{\bf J}\cdot{\bf J}-{\bf L}\cdot{\bf L}
-\frac{1}{4}I\right]\,.   \label{DSQR}
\eea

There are four classes of irreducible representations of the Lie algebra $su(1,1)$.
The first and the second classes include two principal discrete representations, which are realized in the Hilbert space through 
\be
{\cal D}^{\pm}(j)=\left\{\left| j,m_j \right\rangle : \hspace{5mm} j>0, \hspace{5mm} m_j=\pm j,  \pm\left(j+1\right),   \pm\left(j+2\right),\cdots \right\} \label{DPM}
\ee
where
\be
{\bf J}^2 \left| j,m_j \right\rangle=-j(j-1)\left| j,m_j \right\rangle ,  \hspace{25mm} 
J_3 \left| j,m_j \right\rangle=m_j\left| j,m_j \right\rangle \,. 
\label{REPRE}
\ee
It is clear that the state $\left| j,j \right\rangle$ has the lowest weight $j$ in the 
Hilbert space ${\cal D}^{+}(j)$, while the state $\left| j,-j \right\rangle$ has the highest
weight $-j$ in the Hilbert space ${\cal D}^{-}(j)$. Also, there exist principal
continuos representations on the Hilbert space
\be
{\cal C}_{\alpha}(\varsigma)=\left\{\left|\varsigma ,\alpha;m_{\alpha} \right\rangle : \hspace{5mm} \varsigma \in I\!\!R^+ ; \hspace{5mm}  0\leq\alpha<1; \hspace{5mm} m_{\alpha}=\alpha+n ,\hspace{5mm} n\in Z \right\}
\ee
where
\be
{\bf J}^2 \left|\varsigma ,\alpha;m_{\alpha} \right\rangle=\left(\varsigma^2+\frac{1}{4}\right)
\left|\varsigma ,\alpha;m_{\alpha} \right\rangle ,  \hspace{25mm} 
J_3 \left|\varsigma ,\alpha;m_{\alpha} \right\rangle =m_{\alpha} \left|\varsigma ,\alpha;m_{\alpha} \right\rangle .
\ee
Furtheremore, the Lie algebra $su(1,1)$ has complementary continuos representations, which are realized in the Hilbert space as 
\bea
&&{\cal C}_{\alpha}(\tau)=\{\left|\tau ,\alpha;m_{\alpha} \right\rangle : \hspace{5mm} 
-\frac{1}{2}<\tau<-\alpha , \hspace{3mm} 0\leq\alpha<\frac{1}{2}  
\nonumber \\
&&\hspace{30mm} \mbox{or} \hspace{2mm} -\frac{1}{2}<\tau<\alpha-1 , \hspace{3mm} \frac{1}{2}<\alpha<1 ;
\hspace{5mm} m_{\alpha}=\alpha+n , \hspace{5mm} n\in Z \}
\eea
where
\be
{\bf J}^2 \left|\tau ,\alpha;m_{\alpha} \right\rangle=-\tau\left(\tau+1\right)
\left|\tau ,\alpha;m_{\alpha} \right\rangle ,  \hspace{25mm} 
J_3 \left|\tau ,\alpha;m_{\alpha} \right\rangle =m_{\alpha} \left|\tau ,\alpha;m_{\alpha} \right\rangle .
\ee
It is clear that the eigenvalues of the Casimir operator ${\bf J}^2$
corresponding to the principal and complementary continuos representations
are the same if we choose $\tau=-\frac{1}{2}+i\varsigma$ .
It is well known that the decomposition of the tensor product of positive
(or negative) discrete representations of the Lie algebra $su(1,1)$ is
\be
{\cal D}^{\pm}(j)\otimes {\cal D}^{\pm}(s)\cong
\bigoplus_{l\geq j+s }{\cal D}^{\pm}(l)\,. \label{DISCRET}
\ee
And, for $j\geq s$ 
the decomposition of the discrete representations become
\bea
&&{\cal D}^{+}(s)\otimes {\cal D}^-(j)\cong
{\int_0^{\oplus}}^\infty {\cal C}_{s-j+\mu}(\varsigma) d\varsigma
\hspace{25mm} j-s \leq \frac{1}{2} \hspace{4mm} \mbox{and} \hspace{4mm}
j+s\geq \frac{1}{2} \nonumber  \\
&&{\cal D}^{+}(s)\otimes {\cal D}^-(j)\cong
{\int_0^{\oplus}}^\infty {\cal C}_{s-j+\mu}(\varsigma) d\varsigma
\oplus {\cal C}_{s-j+\mu}(\tau=-j-s)
\hspace{17mm} j+s<\frac{1}{2} \nonumber \\
&&{\cal D}^{+}(s)\otimes {\cal D}^-(j)\cong
{\int_0^{\oplus}}^\infty {\cal C}_{s-j+\mu}(\varsigma) d\varsigma
\oplus\bigoplus_{\frac{1}{2}<l\leq j-s}{\cal D}^-(l)
\hspace{28mm} j-s>\frac{1}{2}\, ,   \label{CONTIN}
\eea
where $j-s\leq\mu<j-s+1$, and all the direct sums  are in integer steps.
Suppose the kets $\left| j,m_j \right\rangle$ span the two principal
discrete representation spaces of the operator ${\bf J}$, as given in
(\ref{JLS}) as (\ref{REPRE}). According to the decomposition (\ref{DISCRET}) 
and (\ref{CONTIN}), and since $m_{l}=m_{j}\pm\frac{1}{2}$,
the allowed discrete representations of ${\bf L}^2$ with the
spectrum ${\bf L}^2=-(j+m-\frac{1}{2})(j+m-\frac{3}{2})$ is in 
${\cal D}^{\pm}(l=j+m-\frac{1}{2})$. Note that $m\geq 0$ for both 
possitive and negative principal
discrete representations in (\ref{DISCRET}), and 
$1-j<m\leq -1$ for the negative principal
discrete representation in the last equation of (\ref{CONTIN}). 
If we denote the eigenvalues of $D$ by $\lambda_{j,m}$, 
using Eqs. (\ref{XXXX}),  
the spectrum of the squared Dirac operator is then calculated as
\be
\lambda_{j,m}^2=
\left(j+m-\frac{1}{2}\right)^2\left[1
+\frac{1-\left(j+m-\frac{1}{2}\right)^2}
{N(N-2)}\right]+m(2j+m-1)
\left[\frac{4\left(j+m-1\right)^2-1}
{2N(N-2)}-1\right]\, ,   
\label{SPECTR}
\ee
where the operators ${\bf J}^2$  and ${\bf L}^2$ are restricted over their common
eigenstates. According to the above equation the spectrum of the Dirac operator on fuzzy  $AdS_2$ depends on both quantum numbers $j$ and $m$, whereas in the commutative 
case (hyperbolic plane) the spectrum depends  only on $j$. It is interesting to
see whether the commutative spectrum can be reobtained from the above 
noncommutative expression. In the limit $N\rightarrow \infty$, observe that
$\lambda_{j,m}^2\rightarrow \left(j-\frac{1}{2}\right)^2$ which depends
only on the quantum number $j$, and this is consistent with the fact
that the spectrum of the Dirac operator on $AdS_2$ is
degenerate. Hence, noncommutativity has lifted the degeneracy. Finally note that 
for a finite $N$, the state $j=N-\frac{1}{2}$ and $m=0$ is a zeromode, and 
if we require the spectrum $\lambda_{j,m}$ to be real, we find
that $j+m\leq N-\frac{1}{2}$.

\hspace{20mm}

\hspace{-6mm}{\large \textbf{Acknowledgement}}
\newline
We would like to thank A. Chamseddine for useful correspondence. 
A.I. would also like to thank the theory division of CERN for hospitality 
where part of this work was completed.


\begin{thebibliography}{11}

\bibitem{BAE}
S. Baez, A. P. Balachandran, S. Vaidya and B. Ydri,  
{\em Monopoles and Solitons in Fuzzy Physics}, 
Comm. Math. Phys. {\bf 208} (2000) 787, {hep-th/9811169}.


\bibitem{CON2}
A. Connes and G. Landi, {\em Noncommutative Manifolds the Instanton Algebra and 
Isospectral Deformations}, Commun. Math. Phys. {\bf 221} (2001) 141-159, {math.QA/0011194}. 


\bibitem{CON}
A. Connes, {\em Noncommutative Geometry}, Academic Press, 
London, 1994.

\bibitem{WATA1} U. Carow-Watamura and S. Watamura,
{\em Chirality and Dirac Operator on Noncommutative Sphere},
Comm. Math. Phys. {\bf 183} (1997) 365-382, {hep-th/9605003}.


\bibitem{WATA2} U. Carow-Watamura and S. Watamura,
{\em Noncommutative Geometry and Gauge Theory on Fuzzy Sphere},
Comm. Math. Phys. {\bf 212} (2000) 395-413, {hep-th/9801195}.

\bibitem{HO}
P. Ho and M. Li, {\em Large N Expansion from Fuzzy $AdS_2$}, 
Nucl. Phys. B {\bf 590} (2000) 198-212, {hep-th/0005268}.


\bibitem{FAKH} 
H. Fakhri, {\em Solution of the Dirac Equation 
on the Homogeneous Manifold $SL(2,{\bf C})/GL(1,{\bf C})$ 
in the Presence of a Magnetic Monopole Field},
J. Phys. A: Math. Gen. {\bf 33} (2000) 293-305;
{\em Dirac Equation for a Spin-$\frac{1}{2}$ Charged Particle on the 2D Sphere
$S^2$ and the Hyperbolic Plane $H^2$},
J. Phys. A: Math. Gen. {\bf 35} (2002) 6329-6337.


\bibitem{POLY} B. Morariu and A. P. Polychronakos,
{\em Quantum Mechanics on Noncommutative Riemann Surfaces},
Nucl. Phys. B {\bf 634} (2002) 326-338, {hep-th/0201070}.


\bibitem{BURES}
J. Bures, {\em Dirac Operator on Hypersurfaces},
Comment. Math. Univ. Carolin {\bf 34} (1993) 313-322;
X. Zhang, {\em Lower Bounds for Eigenvalues of Hypersurface Dirac Operators},
Math. Res. Lett. {\bf 5} (1998) 199-210;
X. Zhang, {\em A Remark: Lower Bounds for Eigenvalues of Hypersurface Dirac Operators},
Math. Res. Lett. {\bf 6} (1999) 465-466;
B. Morel, {\em Eigenvalue Estimates for the Dirac-Schr\"odinger Operators},
J. Geom. Phys. {\bf 38} (2001) 1-18.


\bibitem{TRA}
A. Trautman, {\em The Dirac Operator on Hypersurfaces},
Acta Phys. Polon. B {\bf 26} (1995) 1283-1310, {hep-th/9810018}.


\bibitem{MOST} A. Mostafazadeh,
{\em Pseudo-Hermiticity Versus $PT$-Symmetry: The Necessary
Condition for the Reality of the Spectrum of a Non-Hermitian
Hamiltonian}, J. Math. Phys. {\bf 43} (2002) 205-214,
{math-ph/0107001}; {\em Pseudo-Hermiticity Versus $PT$-Symmetry
II: A Complete Characterization of Non-Hermitian Hamiltonians with
a Real Spectrum}, J. Math. Phys. {\bf 43} (2002) 2814-2816,
{math-ph/0110016}; {\em Pseudo-Hermiticity Versus $PT$-Symmetry
III: Equivalence of Pseudo-Hermiticity and the Presence of
Antilinear Symmetries}, J. Math. Phys. {\bf 43} (2002) 3944-3951,
{math-ph/0203005}; {\em Pseudo-Supersymmetric Quantum Mechanics
and Isospectral Pseudo-Hermitian Hamiltonians}, 
Nucl. Phys. B {\bf 640} (2002) 419-434, {math-ph/0203041}.



\bibitem{AZIZ} T. Ya. Azizov and I. S. Iokhvidov,
{\em Linear Operators in Spaces with an Indefinite metric},
Wiley-Interscience, New York (1989); A. Dijksma and H. Langer,
{\em Operator Theory and Ordinary Differential Operators}, in A.
B\"ottcher (ed.) et. al., {\em Lectures on Operator Theory and its
Applications}, Providence, RI: Am. Math. Soc. Fields Institute
Monographs, {\bf 3} (1996) 75.


\bibitem{LANDI}
G. Landi, {\em Deconstructing Monopoles and Instantons}, 
Rev. Math. Phys. {\bf 12} (2000) 1367-1390, {math-ph/9812004}.







\end{thebibliography}
\end{document}